\documentclass[12pt]{article}
\usepackage{amsfonts,amssymb}
\newcommand{\beq}{\begin{eqnarray}}
\newcommand{\eeq}{\end{eqnarray}}
\newcommand{\half}{\frac12}
\newcommand{\bi}{\bibitem}
\newcommand{\wg}{\wedge}
\newcommand{\sbone}{\bar{\sigma}_{\bar{1}}}
\newcommand{\sbtwo}{\bar{\sigma}_{\bar{2}}}

\begin{document}
{}~ \hfill\vbox{\hbox{hep-th/0311084} \hbox{PUPT-2099} }\break

\vskip 1cm

\begin{center}
\Large{Holomorphic D7-Branes and Flavored $\mathcal{N}=1$ Gauge Theories}

\vspace{20mm}

\normalsize{Peter Ouyang}

\vspace{3mm}

\normalsize{\em Joseph Henry Laboratories, Princeton University,}

\vspace{0.2cm}

\normalsize{\em Princeton, New Jersey 08544, USA}
\end{center}

\vspace{10mm}

\begin{abstract}

\medskip

We consider D7-branes in the gauge theory/string theory
correspondence, using a probe approximation.  The D7-branes have four
directions embedded holomorphically in a non-compact Calabi-Yau 3-fold
(which for specificity we take to be the conifold) and their remaining
four directions are parallel to a stack of D3-branes transverse to the
Calabi-Yau space.  The dual gauge theory, which has $\mathcal{N}=1$
supersymmetry, contains quarks which transform in the fundamental
representation of the gauge group, and we identify the interactions of
these quarks in terms of a superpotential.  By activating three-form
fluxes in the gravity background, we obtain a dual gauge theory with a
cascade of Seiberg dualities.  We find a supersymmetric supergravity
solution for the leading backreaction effects of the D7-branes, valid
for large radius.  The cascading theory with flavors
exhibits the interesting phenomenon that the rate of the cascade slows
and can stop as the theory flows to the infrared.

\end{abstract}

\newpage

\section{Introduction}

The gauge theory/string theory correspondence \cite{jthroat,US,EW}
furnishes, in principle, a powerful set of tools for understanding
gauge theories at strong coupling by performing computations in a dual
string theory at weak coupling.  However, the correspondence is only
well-understood in systems where the string background is highly
symmetric and nearly flat, while we expect that the duals to many
interesting gauge theories (such as large-$N$ QCD or SQCD) will not
have these properties.  It is therefore an interesting challenge to
study less symmetric string backgrounds, and in particular to study
backgrounds with reduced supersymmetry.

There is considerable evidence that the string theory dual to the pure
$\mathcal{N}=1$ $SU(N)$ supersymmetric gauge theory in four dimensions
is related, at least in the infrared, to a geometry similar to that of
a warped deformed conifold with flux \cite{KS,MN,Vafa}.  This gauge
theory exhibits chiral symmetry breaking and confinement at low
energies. One interesting generalization of the supersymmetric pure
glue theory is a gauge theory with added flavors.  In the string dual,
the gluon degrees of freedom come from 3-3 strings living on a stack
of D3-branes, while the flavors come from additional open strings
stretching to branes of higher dimension.  These additional branes are
usually D7-branes, but in some setups, the additional branes may be
D5-branes.  In any event, the important feature seems to be that the
added branes must be extended along the radial AdS direction; then,
volume factors suppress the dynamics of the NN strings on these
``flavor branes'', which then contribute states to the gauge theory
with global symmetries rather than gauge symmetries.  The supergravity
dynamics of some related systems were studied in
\cite{afm,granad7,bertd7,Naculich,Karch,Kruczenski,Sakai,Nastase,Wang,Nunez,Babington}.
The papers \cite{Karch,Sakai} studied D7-branes on the conifold, and
\cite{Naculich} studied a similar system with added orientifolds; in
this paper we will continue this study and hopefully provide some new
insights.

One of the ultimate goals of studying D7-branes in AdS
compactifications is to study chiral symmetry breaking in the dual
field theory.  If the added quarks are massless, then the gauge theory
with $K$ flavors possesses a global symmetry $SU(K)\times SU(K)$.  In
QCD, this symmetry is spontaneously broken to its diagonal $SU(K)$
subgroup in the infrared.  We will not reach the goal of finding the
relevant infrared supergravity solution, but we will be able to find a
solution valid at asymptotically large radius (but not so large that
the backreaction of the D7-branes at infinity becomes important; a
similar approximation was used in \cite{granad7,bertd7}.)  We hope to
convince the reader that even this partial solution contains
interesting physics.

The form of the paper is as follows.  In Section 2, we will review the
geometry of the conifold.  We then proceed in Section 3 to add
D3-branes to the conifold, warping it, and to add D7-branes as probes
in the resulting warped geometry.  We identify the corresponding field
theory and study its renormalization group flows for a simple
embedding, and briefly consider some more general D7-brane embeddings.
In Section 4 we add three-form fluxes to the warped conifold, and for
large radius we find an explicitly supersymmetric supergravity
solution including the leading backreaction of the D7-branes.  The
dual field theory exhibits a cascade of Seiberg dualities, for which
the rate of the cascade decreases as the theory runs to low energy;
this behavior appears as a radial dependence of the number of units of
3-form flux in supergravity.  We consider the T-dual of our model in
Section 5 (mostly summarizing work of others) and conclude in Section
6 with several open questions.

\section{Geometry of the Conifold}

In this section we briefly review the geometry of the conifold in
order to fix notation.  Useful references are
\cite{coco,KW,MP,MT,Ceres,Ohta}.

The conifold is a non-compact Calabi-Yau 3-fold, defined by the equation
\beq
z_1 z_2 -z_3 z_4 =0
\label{coneq}
\eeq
in $C^4$. Because Eqn.(\ref{coneq}) is invariant under an overall real
rescaling of the coordinates, this space is a cone, whose base is the
Einstein space $T^{1,1}$ \cite{coco,KW}. The metric on the conifold
may be cast in the form \cite{coco}
\beq
ds_6^2 = dr^2 + r^2 ds_{T^{1,1}}^2\ ,
\label{conimetric}
\eeq
where
\begin{equation} \label{co}
ds_{T^{1,1}}^2=
{1\over 9} \bigg(d\psi +
\sum_{i=1}^2 \cos \theta_i d\phi_i\bigg)^2+
{1\over 6} \sum_{i=1}^2 \left(
d\theta_i^2 + {\rm sin}^2\theta_i d\phi_i^2
 \right)
\ 
\end{equation}
is the metric on $T^{1,1}$. Here $\psi$ is an angular coordinate which
ranges from $0$ to $4\pi$, while $(\theta_1,\phi_1)$ and
$(\theta_2,\phi_2)$ parametrize two ${\bf S}^2$s in a standard way.
This form of the metric shows that $T^{1,1}$ is a $U(1)$ bundle over
${\bf S}^2 \times {\bf S}^2$.

These angular coordinates are related to the $z_i$ variables by
\beq
z_1 &=& r^{3/2} e^{i/2(\psi-\phi_1-\phi_2)}\sin(\theta_1/2)\sin(\theta_2/2), \\
z_2 &=& r^{3/2} e^{i/2(\psi+\phi_1+\phi_2)}\cos(\theta_1/2)\cos(\theta_2/2), \\
z_3 &=& r^{3/2} e^{i/2(\psi+\phi_1-\phi_2)}\cos(\theta_1/2)\sin(\theta_2/2), \\
z_4 &=& r^{3/2} e^{i/2(\psi-\phi_1+\phi_2)}\sin(\theta_1/2)\cos(\theta_2/2).
\label{ztoangles}
\eeq
It is also sometimes helpful to consider a set of ``homogeneous''
coordinates $A_a, B_b$ where $a,b=1,2$, in terms of which the $z_i$
are
\beq
z_1&=&A_1 B_1, \qquad z_2=A_2 B_2 \\
z_3&=&A_1 B_2, \qquad z_4=A_2 B_1. 
\label{ztoAB}
\eeq
With this parameterization the $z_i$ obviously solve the defining equation of the conifold.

We may also parameterize the conifold in terms of an alternative set of complex variables $w_i$, given by 
\beq
\begin{array}{ll}
z_1 = w_1 + iw_2,& \qquad z_2 = w_1 -iw_2 \\
z_3 = -w_3+iw_4,& \qquad z_4 = -w_3 -i w_4.
\end{array}
\label{ztow}
\eeq
The conifold equation may now be written as
\beq
\sum w_i^2 =0
\label{coneqw}
\eeq
and we identify the $T^{1,1}$ base of the cone as the intersection of the conifold with the surface
\beq
\sum |w_i|^2 = r^3.
\eeq
Notice that $T^{1,1}$ described in this way is explicitly invariant
under $SO(4)\simeq SU(2)\times SU(2)$ rotations of the $w_i$
coordinates and under an overall phase rotation.  Thus the symmetry
group of the $T^{1,1}$ is $SU(2)\times SU(2) \times U(1)$.

A useful basis of 1-forms consists of the following holomorphic forms and their complex conjugates:
\beq
\lambda &=& 3\frac{dr}{r} + i \zeta, \\
\sigma_1 &=& \cot(\theta_1/2)(d\theta_1 -i \sin(\theta_1) d\phi_1), \\
\sigma_2 &=& \cot(\theta_2/2)(d\theta_2 -i \sin(\theta_2) d\phi_2),
\label{oneforms}
\eeq
where $\zeta=d\psi+\cos\theta_1 d\phi_1 +\cos\theta_2 d\phi_2$ is the one-form associated with the $U(1)$ fiber of $T^{1,1}$.  A convenient shorthand notation is
\beq
\Omega_{ij} = d\theta_i \wg \sin(\theta_j) d\phi_j.
\eeq
The unusual factors of $\cot\frac{\theta_i}{2}$ appearing in Eq.(\ref{oneforms}) are present so that 
\beq
d\sigma_k= i\Omega_{kk}.
\eeq
To check supersymmetry, we will also need the K\"{a}hler form
\beq
J = -r^2 \left( \frac{1}{3} \frac{dr}{r} \wg \zeta- \frac{1}{6}(\Omega_{11}+\Omega_{22}) \right).
\eeq
Finally, for reasons that will become clear, we will often use the relation
\beq
\frac{dz_1}{z_1}=\half(\lambda+\sigma_1+\sigma_2).
\eeq

\section{Embedding Flavor Branes in $AdS_5 \times T^{1,1}$}

We begin our study of D7-branes by attempting to embed them in the
model of Klebanov and Witten \cite{KW}.  This model is a particularly
simple $N=1$ gauge/gravity dual, obtained by placing a stack of $N$
D3-branes near a conifold singularity.  The branes source the RR
5-form flux and warp the geometry:
\beq
ds_{10}^2 &=& h(r)^{-1/2} (dx_{\mu} dx^{\mu} +dr^2) + h(r)^{1/2}r^2 ds_{T^{1,1}}^2 \\
h(r) &=& \frac{L^4}{r^4}\\
g_s F_5 &=& d^4x \wg dh^{-1} + \star ( d^4x \wg dh^{-1})\\
L^4 &=& \frac{27}{4} \pi g_s N \alpha'^2.
\eeq
The dual field theory has gauge group $SU(N) \times SU(N)$ and matter
fields $A_{1,2},B_{1,2}$ which transform in the bifundamental
color representations $({\bf N},{\bf \bar{N}})_c$ and $({\bf \bar{N}},{\bf
N})_c$.  The theory also has a superpotential
\beq
W=\lambda Tr(A_i B_j A_k B_l) \epsilon^{ik} \epsilon^{jl}.
\eeq
By solving the F-term equations for this superpotential, we can see
that we obtain supersymmetric vacua for arbitrary diagonal $A_{1,2}$
and $B_{1,2}$, so that the moduli space of the field theory is
precisely that of $N$ D3-branes placed on a conifold.

Now we would like to modify this field theory by the inclusion of
fundamental matter, which should correspond on the dual string theory
side to the addition of D7-branes.  These D7-branes fill the 4
dimensions tangent to the D3-branes and must also wrap 4 dimensions in
the conifold.  It is natural to suppose that we will obtain a
supersymmetric solution if the equation specifying the embedding is
holomorphic \cite{Karch} -- then the submanifold corresponding to the
D7-brane worldvolume inherits a complex structure and a closed
K\"{a}hler form from the original Calabi-Yau space (and should
therefore inherit some fraction of the original supersymmetry.  See,
for example, \cite{bbs,ooy}.)  Let us start then with the simple
holomorphic equation $z_1=0$, where $z_1$ is one of the complex
variables in the defining equation of the conifold
(\ref{coneq})\footnote{Our embedding equation is in some sense
``one-half'' of the embedding of \cite{Karch}, who embedded D7-branes
by the equation $z_1 z_2=0$, in our coordinates.  It differs also from
the embedding of \cite{Sakai}, which is not holomorphic, even in the
limit where the RR 3-form flux is turned off.}.  Note that in terms of
the homogeneous coordinates $A,B$, there are two branches of our
D7-brane, $A_1=0$ and $B_1=0$.  There is an $SU(K)$ flavor symmetry
associated with each branch of a stack of $K$ D7-branes, so we expect
the existence of a global $SU(K) \times SU(K)$ flavor symmetry.
Moreover, cancellation of gauge anomalies requires that we add two
flavors to each gauge group, with opposite chiralities.  We denote the
resulting four sets of flavors as $q, \tilde{q},Q,\tilde{Q}$, and
indicate their color and flavor representations in Table 1.
\begin{table}
\begin{center}
\begin{tabular}{||c|c|c||} \hline
Field& $SU(N_c)\times SU(N_c)$& $SU(K)\times SU(K)$ \\ \hline \hline
$q$& $({\bf N},{\bf 1})$& $({\bf K},{\bf 1})$\\ \hline
$\tilde{q}$& $({\bf \bar{N}},{\bf 1})$&$({\bf 1},{\bf K})$\\ \hline
$Q$& $({\bf 1},{\bf N})$& $({\bf \bar{K}},{\bf 1})$\\ \hline
$\tilde{Q}$& $({\bf 1},{\bf \bar{N}})$& $({\bf 1},{\bf \bar{K}})$\\ \hline
\end{tabular}
\end{center}
\caption{Representation structure of the added $\mathcal{N}=1$ flavors.}
\end{table}
We propose that the corresponding gauge invariant and flavor invariant
terms in the superpotential are
\beq
W_{flavors}= h qA_1 Q + g \tilde{q}B_1 \tilde{Q}.
\label{wflavors}
\eeq
If the number of flavors $K$ is much smaller than the size of the
gauge group $N$, then the dimensions of the flavor superfields are
determined (to leading order in $1/N$) by the following argument.
Because the theory including D7-brane probes is invariant under the
rescaling $z_i \rightarrow \beta z_i$, the field theory should be
classically scale invariant.  In the conformal theory without any
flavors, the $A$ and $B$ fields have dimension 3/4.  Therefore, power
counting in the superpotential requires that the $q, Q$ superfields
each have dimension 9/8 (plus quantum corrections.)

It is worth noting that the D7-branes embedded by a holomorphic
equation, as we have described here, are topologically trivial.  This
topological triviality is essential for RR charge conservation; a
topologically nontrivial wrapping would necessitate the presence of
anti-D7-branes or orientifold planes.

It is straightforward to add masses to the flavors, at the cost of
breaking the $SU(K)\times SU(K)$ flavor symmetry to its diagonal
$SU(K)$ subgroup.  The relevant terms in the superpotential are
\beq
W_{masses}=\mu_1 q\tilde{q} +\mu_2Q\tilde{Q}.
\label{wmasses}
\eeq
To translate these masses to the D7-brane probe picture, it is helpful
to rewrite $W_{flavors}+W_{masses}$ in the following matrix form:
\beq
W_{flavors}+W_{masses}=\left(\begin{array}{cc} \tilde{q} & \tilde{Q} \end{array} \right) 
\left(
\begin{array}{cc}
\mu_1& h A_1  \\
g B_1& \mu_2  \end{array} \right)
\left( \begin{array}{c}
q\\ Q\end{array} \right).
\label{wmatrix}
\eeq
A useful technique for relating the string theory and field theory is
to probe the string background with D-branes; for simplicity, let us
consider a single D3-brane probe, so that the fields $A_1$ and $B_1$
are just scalars.  Giving $A,B$ vacuum expectation values, we can
think of the square matrix in (\ref{wmatrix}) as a mass matrix for the
quarks.  When the D3-brane and D7-brane intersect, some of the quarks,
which arise as 3-7 strings, become massless.  In other words, the
determinant of the mass matrix $h g A_1 B_1 -\mu_1 \mu_2$ should
vanish when the D3-brane probe is on the D7-brane locus, or with an
appropriate redefinition,
\beq
z_1 - \mu^2=0.
\label{z1mu2}
\eeq
Thus Eq.(\ref{z1mu2}) is also an appropriate embedding equation for a
D7-brane which gives massive flavors (when the D3-branes are located
at the tip of the cone.)

To obtain further evidence for the validity of our construction, let
us study the axion and dilaton fields sourced by our D7-brane; in
particular, we can obtain RG flows in the field theory from the
running of the dilaton in supergravity.  It is convenient to consider
the axion $C_0$ and dilaton $\Phi$ in the complex combination $\tau =
C_0 +ie^{-\Phi}$.  Given the embedding condition $z_1=0$, a natural
guess for the combined dilaton-axion system is that
\beq
\tau \sim \log(z_1).
\label{logz1}
\eeq
With the normalization condition
\beq
\int_{S^1} F^1 = N_{D7} =K
\label{fonenorm}
\eeq
we see that
\beq
C_0 = \frac{K}{4\pi} \left(\psi-\phi_1-\phi_2 \right).
\label{czero}
\eeq
Thus we have
\beq
\tau = \frac{i}{g_s}+\frac{K}{2\pi i} \log z_1,
\label{taunorm}
\eeq
which gives the correct $SL(2,Z)$ monodromy $\tau \rightarrow \tau+K$
upon circling a stack of $K$ D7-branes; the first term has been chosen
to give the correct value of the dilaton when $K=0$.  When $K\neq 0$,
Eq.(\ref{taunorm}) shows that the dilaton is given by
\beq
e^{-\Phi} = \frac{1}{g_s} - \frac{3K}{4 \pi} \log r -\frac{K}{2\pi}\log\left(\sin\frac{\theta_1}{2}\sin\frac{\theta_2}{2}\right).
\eeq
The holomorphic dilaton-axion described in Eq.(\ref{taunorm}) is
manifestly a supersymmetric solution of the supergravity equations of
motion with the three-form fluxes set to zero.  In the probe
approximation we are using, we ignore the backreaction on the geometry
and RR five-form.  The singularity of (\ref{taunorm}) is acceptable
for our purposes, as it corresponds to the presence of a D7-brane.  As
is usually the case with D7-branes, for small enough $z_1$ the dilaton
may become negative; it would be interesting to find a solution using
F-theory \cite{GSVY} that avoids this problem.

Recall that the couplings for the two gauge groups are determined as
follows\footnote{Strictly speaking, these formulae have been derived
for the case of branes at orbifold singularities and the corresponding
gauge theories.  For systems such as our conifold theory, there is no
proof from first principles, but the RG flows have been checked in
several cases.} \cite{KW,MP}:
\beq
\label{rg1}
{4\pi^2 \over g_1^2} + {4\pi^2 \over g_2^2} ={\pi\over e^\Phi}
\ ,
\eeq
\beq \label{rg2}
\left [ {4\pi^2 \over g_1^2} - {4\pi^2 \over g_2^2}\right ]
e^\Phi
= {1\over 2\pi \alpha'}
\left(\int_{S^2} B_2\right) - \pi \quad ({\rm mod}\ 2\pi)
\ .
\eeq
In this theory with no three-form fluxes in supergravity, we may set
$g_1=g_2=g_{YM}$.  Making the identification of the AdS radius $r$ as
the renormalization scale $\Lambda$ \cite{jthroat}, we find that
\beq
\frac{\partial}{\partial \log\Lambda} \frac{8 \pi^2}{g_{YM}^2} = -\frac{3K}{4}.
\label{rgsugra}
\eeq

Let us compare this RG equation with the Shifman-Vainshtein $\beta$-functions \cite{SV,NSVZ}:
\beq
\frac{\partial}{\partial \log\Lambda} \frac{8 \pi^2}{g_{YM}^2} = 3N - 2N (1-2\gamma_{A,B}) - K (1-2\gamma_{q}).
\label{svbeta}
\eeq
Each gauge group has $2N$ effective flavors coming from the $A,B$
bifundamentals and $K$ effective flavors from half of the $q,Q$
fields.  With the anomalous dimensions $\gamma_{A,B} = -1/4$, the
terms proportional to $N$ cancel, as necessary for conformal
invariance of the background.  However, we argued earlier that the
dimension of the $q$ fields is 9/8, corresponding to $\gamma_q=1/8$.
With this assignment we see that the RG flows from supergravity and
field theory agree precisely at this order in the $1/N$ expansion,
provided that the anomalous dimensions of the $A$, $B$ bifundamental
fields do not receive corrections of order $K/N$.  It would be nice to
check explicitly that such corrections do not appear. one indication
that this is the case comes from supergravity -- from Einstein's
equations one sees that the metric receives leading backreaction
corrections of order $K^2/N^2$, so at first order in $K/N$ the
spacetime geometry is anti-de Sitter.  Therefore we should expect the
field theory to be conformal including terms of order $K/N$, as we
have assumed.  The RG flows are related by supersymmetry to the
$\theta$-angles of the gauge theory.  This relationship was studied on
the conifold in \cite{kow}; more detailed supergravity analysis
appears in \cite{Krasnitz,BDFP,brand}.

There is a second possibility for the superpotential that will produce
the same D7-brane geometry (the holomorphic embedding $z_1=\mu^2$) from
the probe D3-brane perspective -- consider the addition of only two
flavors, $q$ and $Q$, with the superpotential
\beq
W_{flavors}= q(A_1 B_1 +\mu_1)\tilde{q}.
\label{qABQ}
\eeq
However, this superpotential actually arises as a special case of the
cubic superpotential (\ref{wflavors}) considered earlier, by adding a
mass term of the form (\ref{wmasses}).  Integrating out the the
$Q, \tilde{Q}$ flavors, one obtains the quartic superpotential
(\ref{qABQ}).

Another way to motivate the gauge theory construction presented here
is to consider a related $C^2/Z_2$ orbifold theory, where standard
arguments \cite{Douglas} allow us to construct the field theory
explicitly.  Then we can obtain the conifold field theory by turning
on a particular mass deformation \cite{KW}.  The relevant orbifold background may
be defined by starting in flat ten-dimensional space and then
orbifolding by the discrete symmetry
\beq
X^{6,7,8,9} \rightarrow -X^{6,7,8,9}.
\label{z2orb}
\eeq
An equivalent definition is to start with a $C^3$ space parametrized
by the coordinates $z_1,z_2,z_3$ and to then consider the submanifold
defined by the equation
\beq
z_1 z_2 - z_3^2 = 0.
\label{z2equation}
\eeq
The resulting four-manifold, tensored with six-dimensional flat space,
is the same as the orbifold defined by Eq.(\ref{z2orb}).  We may
``solve'' Eq.(\ref{z2equation}) by introducing complex variables $A,B$
such that
\beq
z_1 = A^2, \qquad z_2 = B^2, \qquad z_3 = AB.
\eeq
The invariance of the $z_i$ coordinates under $A,B\rightarrow -A,-B$
expresses the $Z_2$ orbifold action.

By placing $N$ D3-branes transverse to this orbifold, we will obtain
an $SU(N) \times SU(N)$ SCFT with $N=2$ supersymmetry, and
bifundamental matter superfields $A_{1,2}$ and $B_{1,2}$ (there must
be two of each because of the $N=2$ supersymmetry.)  The quiver
diagram for this theory consists of two nodes, corresponding to a
stack of D3-branes and its mirror image under the orbifold action.
The vector multiplets arise as strings connecting each node to itself,
while the matter fields arise as strings connecting different nodes.
It was shown in \cite{KW} that if one deforms this theory by giving
masses to the adjoint scalars in the $N=2$ vector multiplets, and
subsequently integrates out the adjoints, the theory becomes the $N=1$
conifold theory described earlier.

To add flavors, we now consider the addition of D7-branes, with 4
directions tangent to the D3-branes.  One simple way to embed these
branes is to wrap the remaining 4 directions on the entire
orbifold\cite{granad7,bertd7}.  An alternative is to consider a
holomorphic embedding equation (such as $z_1=0$) so that the D7 fills
two dimensions of the orbifold and the remaining six spacetime
directions.  Because the homogeneous coordinates that parametrize the
orbifold are double-valued, including a D7-brane in this way can
actually be thought of as the inclusion of two fractional D7-branes --
a D7 and its orbifold mirror image (another way to think about this is
to deform the defining equation of the orbifold to $z_1 z_2 =
z_3^2-\varepsilon^2$.  Then $z_1=0$ has two branches, $z_3=\pm
\varepsilon$.)  It is then clear from considering the quiver diagram
that we should obtain superpotential interactions of the form of
Eq.(\ref{wflavors}), and that these terms survive when we integrate
out the adjoints to arrive at the $N=1$ theory of interest.

The arguments in this section should be more or less unchanged for the
case of D7-branes in the ``generalized conifold'' geometries discussed
in \cite{gns}.  In particular, the dimensions of the bifundamental
matter fields are still 3/4, so the flavors should still have
dimension 9/8, and the RG flow equations should work out the same way.
It might be interesting to study conifolds in other dimensions
\cite{hk,cveticmetricfactory}.

\subsection{Other Embedding Equations}

Though we have obtained sensible results for the simple embedding
equation $z_1=\mu^2$, it is clearly interesting to consider a more
general polynomial $P(z_i)$.  In this section we present some simple
considerations on how one might realize such D7-brane embeddings might
be realized in the dual field theory.

The simplest generalization of the D7-branes considered thus far
arises from the symmetries of the conifold geometry.  The base of the
conifold possesses a geometric $SO(4)$ symmetry which is spontaneously
broken by the probe D7-brane; therefore we may perform such rotations
to obtain alternative brane embeddings with the same shape but
different positions in the conifold (the remaining geometric $U(1)$
symmetry corresponds to multiplication of the mass parameter $\mu$ by
a phase.)  These rotations are simple to describe in terms of the
$w_i$ coordinates of (\ref{ztow}), which transform simply under
$SO(4)$.  The embedding equation $z_1=\mu^2$ becomes $w_1+iw_2=\mu^2$,
and under a general $SO(4)$ transformation this becomes
\beq
\sum_{j=1}^{4}a_j w_j + i b_j w_j=\mu^2
\eeq
with the constraint that 
\beq
\sum a_j b_j=0, \qquad \sum a_j^2=\sum
b_j^2,
\label{so4constr}
\eeq
for real $a_j, b_j$.  Because all these embeddings are defined
holomorphically, with respect to the same complex structure, we 
expect that they preserve supersymmetry.  

Turning to the field theory, we also find that these rotated D7-branes
are easy to describe with $SO(4)$ realized as the $SU(2)\times SU(2)$
action on the matter fields $A_a,B_a$.  The relevant group action now
sends $hA_1$ and $gB_2$ in the superpotential (\ref{wflavors}) to
$h_1A_1+h_2A_2$ and $g_1 B_1+g_2B_2$, respectively.

However, there are many holomorphic embeddings which do not satisfy
the constraints (\ref{so4constr}); it is natural to ask if these other
embeddings are allowed as well.  For specificity, let us consider the
explicit example $w_1=z_1+z_2=0$ (this equation was studied in \cite{Naculich} with the addition of orientifold planes.)  Though we cannot realize this
embedding as the determinant of a $2\times 2$ matrix of the form in
(\ref{wmatrix}) we can obtain it from a $4\times 4$ matrix:
\beq
\left(
\begin{array}{cccc}
0& 0& A_1 & A_2 \\
0& 1& 0& 0 \\
B_1& 0& 1 &0 \\
B_2& 0& 0&1\end{array} \right)
\label{w1matrix}
\eeq
which (heuristically) corresponds to having 2 D7-branes, whose 7-7
strings receive expectation values to give the necessary coupling
constants in the field theory.

We can also realize this embedding equation $w_1=0$ by considering
the related $C^2/Z_2$ orbifold theory with D3-branes transverse to the
orbifold.  For this theory it has been shown that one can embed
D7-branes by wrapping four worldvolume dimensions on the entire
orbifold, and placing the other 4 dimensions parallel to the
D3-branes.  In the field theory, the quarks that arise are coupled to
the adjoint scalar field $\Phi$, breaking the supersymmetry from
$\mathcal{N}=2$ to $\mathcal{N}=1$ (there are actually two such
embeddings, corresponding to a choice of orientation in the string
background, and corresponding to a choice of which of the two adjoint
scalars $\Phi$ and $\tilde{\Phi}$ to which the quarks are coupled in
the field theory.)  The superpotential is
\beq
W= g {\rm Tr} \left((A_1 B_1 + A_2 B_2)\Phi\right)+g {\rm Tr} \left((B_1 A_1 + B_2 A_2)\tilde{\Phi}\right) +g q\Phi Q.
\eeq
Now, by giving masses to the adjoints,
\beq
W \rightarrow W+ \frac{m}{2}\Phi^2-\frac{m}{2}\tilde{\Phi}^2
\eeq
and integrating them out, we obtain a superpotential 
\beq
W=Tr(A_i B_j A_k B_l) \epsilon^{ik} \epsilon^{jl} - (q\tilde{q})^2-2q(A_1 B_1 + A_2 B_2)\tilde{q}.
\label{quarticq}
\eeq
The locus on which there are massless flavors is clearly $w_1=0$.
Also, because the embedding equation is still scale invariant, we
should expect the theory to be classically conformal.  Thus the
flavors all have dimension 3/4.  The Shifman-Vainshtein
$\beta$-functions with this assignment agree with supergravity if, in
addition to a nontrivial dilaton, there are some 2-form potentials
turned on.  Note that the alternative quartic superpotential
(\ref{quarticq}) does not actually represent new physics, as it can be
obtained from the cubic superpotential defined by (\ref{w1matrix}) by
integrating out a set of flavors, and adding the interaction
$(q\tilde{q})^2$, which becomes marginal in the IR.  It would be nice
to find the explicit supergravity solution for this D7-brane embedding
on the conifold, and also to determine whether the cubic interactions encoded in (\ref{w1matrix}) can be realized directly from string theory.

It is also possible to consider higher-order polynomials in the $z_i$.
If the polynomial factors into lower-order polynomials, then the
separate factors have a natural interpretation as disjoint D7-branes.
On the other hand, if the polynomial does not factor, then the picture
changes slightly.  For example, let us consider the polynomial
$z_1z_2-\mu^4=0$.  Once again, we may realize this embedding equation
as the determinant of a mass matrix:
\beq
\left(
\begin{array}{cccc}
0& \mu& A_1 & 0\\
\mu& 0& 0& A_2\\
B_1& 0& \mu &0 \\
0& B_2& 0&\mu\end{array} \right)
\eeq
At high energies, or equivalently at large $AdS$ radius, $z_1$ and
$z_2$ are large compared to the mass scale $\mu^2$ and so the theory
seems to have eight quark superfields.  At low energies, the mass
perturbations allow us to integrate out some of the flavors.  The
analogous string picture is that we start with two D7-branes embedded
by $z_1=0$ and $z_2=0$, which intersect at the tip of the conifold. By
turning on appropriate 7-7 strings, the D7-branes fuse near the
intersection.

\section{Fractional D3-branes and Flavored Cascades}

We can obtain an interesting generalization of the conifold theory
described in the previous section by introducing three-form fluxes on
the conifold.  These models were studied in a series of papers
\cite{KN,KT,KS} and were shown to be dual to an $N=1$ theory with
gauge group $SU(N+M)\times SU(N)$, where $N$ is the number of units of
5-form flux and $M$ is the number of units of 3-form flux in the
background.  The 3-form flux is sourced by D5-branes which are wrapped
on the $S^2$ in $T^{1,1}$; in the literature these wrapped D5-branes
are often called ``fractional D3-branes.''

In the large radius limit, the supergravity solution including these
3-form fluxes (but with the axion and dilaton constant) is
\beq
ds_{10}^2 &=& h(r)^{-1/2} (dx_{\mu} dx^{\mu} +dr^2) + h(r)^{1/2}r^2 ds_{T^{1,1}}^2 \label{ktmetric}\\
h(r) &=& \frac{27}{4r^4} \pi g_s\alpha'^2 \left( N + \frac{3}{2\pi} g_s M^2 \log(r/r_0)\right)\label{ktwarp}\\
g_s \tilde{F}_5 &=& d^4x \wg dh^{-1} + \star ( d^4x \wg dh^{-1})\label{ktf5}\\
F_3 &=& \frac{M\alpha'}{2}\omega_3 \label{ktf3}\\
H_3 &=& \frac{3g_s M\alpha'}{2} \frac{dr}{r} \wg \omega_2. \label{kth3}
\eeq
This solution possesses a naked singularity at small $r$; a
nonsingular solution was found in \cite{KS} by deforming the conifold.
Though it would certainly be interesting to study D7-branes on a
deformed conifold, we will not attempt to do so here.

Another feature of the supergravity solution with fractional branes,
which will be quite important for this paper, is the logarithm in the
warp factor $h(r)$, and the corresponding logarithmic running of the
5-form flux $\tilde{F}_5$.  The number of units of 5-form flux is dual
to the number of colors in the field theory, so the radial dependence
of $\tilde{F}_5$ implies that the gauge groups effectively decrease in
size as the theory undergoes renormalization group flow from the
ultraviolet to the infrared:
\beq
N_{eff} =  N + \frac{3}{2\pi} g_s M^2 \log(r/r_0).
\eeq
This phenomenon arises from a ``cascade'' of Seiberg dualities, which
we review briefly here, without dwelling on technical details.
Suppose that we begin with a gauge theory with gauge group
$SU(N+M)\times SU(N)$.  The $SU(N+M)$ factor is coupled to $2N$
effective flavors, while the $SU(N)$ factor couples to $2(N+M)$
effective flavors.  From the NSVZ $\beta$-functions, it is clear that
the $SU(N+M)$ factor of the gauge group flows toward strong coupling
in the infrared, while the $SU(N)$ factor flows to weak coupling.
When the $SU(N+M)$ factor becomes strongly coupled, we may perform a
Seiberg duality transformation, under which a strongly coupled gauge
theory with gauge group $SU(N_c)$ and $N_f$ flavors becomes a weakly
coupled theory with gauge group $SU(N_f-N_c)$ and $N_f$ flavors (there
are also mesonic states, which acquire mass and are irrelevant to our
discussion.)  In the present case, the duality sends the first gauge
group to $SU(2N-(N+M))=SU(N-M)$.  After the duality transformation, we
are left with an $SU(N)\times SU(N-M)$ gauge theory with $A,B$
bifundamental flavors; in other words, the number of colors $N$ has
effectively decreased by $M$.  Because the process just described will
now repeat until $N\sim M$, it is known as a ``cascade.''

We will again attempt to find a supergravity solution corresponding to
a D7-brane with the embedding $z_1=0$, using the axion-dilaton ansatz
found in the previous section.  Once again, for our purposes the only
requirements are that the axion-dilaton is holomorphic and that it has
the correct monodromy.  What is different in this case is that the
background three-form fluxes will induce three-form charges in the
D7-branes; we will compute the leading order corrections to the
fluxes, but treat the branes themselves as probes.  

The addition of fundamental flavors has a pronounced effect on the
pattern of duality cascades, which we will attempt to reproduce in
this section from supergravity.  With $K$ D7-branes, the number of
effective flavors coupled to each gauge group increases by $K$.  If we
start with a gauge group $SU(N+M)\times SU(N)$, then because the first
gauge group factor has $2N+K$ effective flavors, it is Seiberg-dual in
the infrared to an $SU(2N+K-(N+M))=SU(N-M+K)$ gauge theory.  The
second gauge group factor remains $SU(N)$, so we see that in addition
to a decrease of the overall number of colors, the {\em difference} in
the size of the gauge groups has decreased from $M$ to $M-K$.  As we
continue to follow the renormalization group, we see that at each step
of the duality cascade, the strength of the cascade, $M$, decreases by
$K$.  A very similar phenomenon was described in the paper
\cite{Franco}, who performed a field theory analysis of del Pezzo
theories with bifundamental matter; it would be interesting to
understand the stringy description of their duality cascade.

The difference in the size of the gauge groups, $M$, will decrease by
increments of $K$ until it is smaller than or equal to $K$.  If at
this point $N$ is still greater than zero, we should have the
$SU(N)\times SU(N)$ theory with $K$ flavors considered in Section 3.
An alternative possibility is that $N$ may decrease to zero, but with
a finite $M$ left over.  Then we would expect the field theory to be
that of $\mathcal{N}=1$ $SU(M)$ SYM with $K$ flavors.

\subsection{Three-form Fluxes and RG Flow}

We now turn to the three-form fields $F_3$ and $H_3$.  It is
convenient to consider these fields in the complex combination $G_3 =
F_3 -\tau H_3$. The supersymmetric solutions that we will study
satisfy the conditions that $G_3$ is imaginary self-dual, has index
structure (2,1), and is primitive (contractions with the K\"{a}hler
form vanish), that the three-forms and five-form satisfy Bianchi
identities, and that the metric is a ``warped product.''
\cite{Grana,Gub}

To take advantage of the simplifications made possible by
supersymmetry, we introduce a basis of imaginary self-dual (2,1) forms
for the conifold:
\beq
\eta_1 &=& \lambda \wg \omega_2 \\
\eta_2 &=& \half \lambda \wg (\sigma_1\wg\sbtwo -\sigma_2 \wg \sbone) \nonumber \\
&=& \cot(\theta_1/2)\cot(\theta_2/2) \lambda \wg (d\theta_1 \wg d\theta_2 +\sin(\theta_1) d\phi_1 \wg \sin(\theta_2) d\phi_2 )\\
\eta_3 &=&\left(\frac{dr}{r} \wg \zeta + \half \Omega_{22} \right) \wg \sigma_1, \\
\eta_4 &=&\left(\frac{dr}{r} \wg \zeta + \half \Omega_{11} \right) \wg \sigma_2, \\
\eta_5 &=& \bar{\lambda} \wg \sigma_1 \wg \sigma_2 \nonumber \\
&=& \bar{\lambda} \wg (d\theta_1 \wg d\theta_2 -\sin(\theta_1) d\phi_1 \wg \sin(\theta_2) d\phi_2 - i (\Omega_{12}-\Omega_{21})).
\eeq
It is a happy coincidence that all these (2,1) forms are primitive as well.  In this basis, the background $G_3$ is just
\beq
G_3^{(0)} = F_3^{(0)} -\frac{i}{g_s} H_3^{(0)} = -i \frac{M\alpha'}{2} \eta_1.
\eeq
Under exterior differentiation, these 3-forms give
\beq
d(\eta_1) &=& 0 \\
d(\eta_2) &=& \frac{i}{2} \lambda \wg (\Omega_{22} \wg (\sbone-\sigma_1)-\Omega_{11}\wg(\sbtwo-\sigma_2))\\
d(\eta_3) &=& \frac{dr}{r} \wg \Omega_{22}\wg \sigma_1+i \frac{dr}{r}\wg \zeta\wg\Omega_{11}-\frac{i}{2}\Omega_{11}\wg \Omega_{22} \\
d(\eta_4) &=& \frac{dr}{r} \wg \Omega_{11}\wg \sigma_2+i \frac{dr}{r}\wg \zeta\wg\Omega_{22}-\frac{i}{2}\Omega_{11}\wg \Omega_{22} \\
d(\eta_5) &=& -2i\bar{\lambda} \wg \omega_2 \wg(\sigma_1 +\sigma_2)
\eeq

The three-form $G_3$ satisfies a Bianchi identity because the fields
$F_3$ and $H_3$ are derived from potentials:
\beq
dG_3 = -d\tau \wg H_3 = -\frac{3g_s MK\alpha'}{4\pi i} \frac{dz_1}{z_1} \wg \frac{dr}{r} \wg \omega_2 + O((g_sK)^2).
\label{bianchig3}
\eeq
Notice that the background three-form and the terms we will need to
add to satisfy the Bianchi identity are antisymmetric under the
exchange of the two two-spheres.  This suggests that the forms
$\eta_3$ and $\eta_4$ should appear only in the combination
$\eta_3-\eta_4$.  A three-form satisfying (\ref{bianchig3}) is
\beq
P_3 = \frac{3g_s MK\alpha'}{16\pi i}(4\log(r) \eta_1 -\eta_3+\eta_4).
\label{p3}
\eeq
There is
also a closed 3-form which satisfies the supersymmetry conditions and
has an acceptable singularity structure (logarithmic singularities at
$r=0$ and $z_1=0$):
\beq
Q_3 = \eta_3-\eta_4 - (2\log(r) +\frac23 \log(z_1))\eta_1 +\frac{i}{6} \eta_5.
\label{q3}
\eeq
The expression (\ref{p3}) is a particular solution of the Bianchi
identity, while (\ref{q3}) is a homogeneous solution.  To identify the
linear combination of $P_3$ and $Q_3$ that gives the physical
solution, we should look at the expected singularity structure of
$G_3$ in the presence of D7-branes.  Under the shift $\psi \rightarrow
\psi +4\pi$, the complex scalar transforms as $\tau \rightarrow \tau +
K$, corresponding to the usual $SL(2,Z)$ monodromy around $K$
D7-branes.  However, under this monodromy the complex three-form $G_3$
does not transform (for a general $SL(2,Z)$ transformation, $\tau
\rightarrow \frac{a\tau+b}{c\tau+d}$ and $G_3 \rightarrow
\frac{G_3}{c\tau+d}$.)  Thus the leading correction to $G_3$ is given
by Eq.(\ref{p3}), with no additional contribution from the homogeneous
solution (\ref{q3}):
\beq
G_3= \frac{M\alpha'}{2i} \left[\left(1+\frac{3g_s K}{2\pi}\log r\right)\eta_1 +
\frac{3g_s K}{8\pi}\left(\eta_4-\eta_3\right) \right].
\label{g3}
\eeq

Having determined the three-form fluxes from supergravity, let us now
investigate their effect in the dual field theory.  The RR 3-form flux
$\tilde{F}_3=F_3-C_0H_3=\frac{G_3+\bar{G_3}}{2}$ is
\beq
\hspace{-0.5in}\tilde{F}_3&=& \frac{M\alpha'}{2}\Bigg[\Big(1+\frac{3g_sK}{2\pi}\log r\Big)\zeta\wg \omega_2 \nonumber\\&&+ \frac{3g_sK}{8\pi}\Big(\frac{dr}{r}\wg \zeta+\half(\Omega_{11}+\Omega_{22})\Big) \wg\Big( (1+\cos\theta_1)d\phi_1-(1+\cos\theta_2)d\phi_2\Big)\Bigg]
\eeq
Integrating this flux over the topologically nontrivial 3-cycle of
$T^{1,1}$, $\omega_3=\zeta\wg \omega_2$, we find that the number of
units of flux varies logarithmically as a function of the radius $r$:
\beq
M_{eff}(r)= M(1+\frac{3g_sK}{2\pi}\log r).
\label{meff}
\eeq

This logarithmic running, which is a central result of this paper, is
quite similar to the running of the five-form flux (\ref{ktf5}) in the
background with no D7-branes.  Now, in addition to the variation of
the number of colors, we see that the nontrivial axion and dilaton
cause the {\em difference} in the size of the gauge group factors to
decrease as we follow the RG flow from the ultraviolet to the
infrared.  Moreover, the rate of the decrease agrees with field theory
expectations: for $r \rightarrow re^{-\frac{2\pi}{3g_sM}}$, $N_{eff}$
decreases by $M$, while $M_{eff}$ decreases by $K$.

Recall that the equations (\ref{rg1}) and (\ref{rg2}) relate the
dilaton and $B$ field to the field-theoretic renormalization group
flow.  The NS-NS 3-form flux $H_3$ is given in terms of the complex 3-form
$G_3$ by
\beq
H_3 = \frac{\bar{G}_3-G_3}{\tau-\bar{\tau}}.
\eeq
Working to leading order in $g_s K$, one finds after some algebra that
\beq
H_3 &\simeq& \frac{3g_s M\alpha'}{2}\Bigg[\Big(1+\frac{9g_sK}{4\pi}\log r+\frac{g_sK}{2\pi} \log(\sin\frac{\theta_1}{2}\sin\frac{\theta_2}{2})\Big)\frac{dr}{r}\wg\omega_2 \nonumber \\&&+ \,\frac{g_sK}{8\pi}\Big(\frac{dr}{r}\wg \zeta -\half d\zeta\Big) \wg (\cot\frac{\theta_2}{2}d\theta_2-\cot\frac{\theta_1}{2}d\theta_1) \Bigg].
\eeq
The corresponding two-form potential $B_2$ is given by 
\beq 
B_2&=&\frac{3g_s M\alpha'}{2}\Bigg[\Big(\log r +
\frac{9g_sK}{8\pi}\log^2 r +\frac{g_sK}{4\pi}(1+2\log r)\log(\sin\frac{\theta_1}{2}\sin\frac{\theta_2}{2})\Big) \omega_2 \nonumber \\
&&\qquad +\frac{g_sK\log r}{8\pi} \zeta \wg (\cot\frac{\theta_2}{2}d\theta_2-\cot\frac{\theta_1}{2}d\theta_1) \Bigg],
\eeq
and the form relevant for the RG flow equation (\ref{rg2}) is
\beq
e^{-\Phi}B_2 = \frac{3M\alpha'}{2}\left(\log r +\frac{3g_sK}{8 \pi}\log^2 r\right) \omega_2 + ...
\eeq
where the ellipsis denotes terms that do not affect the RG flow at
linear order in $g_sK$.  Thus
\beq
\frac{\partial}{\partial\log\Lambda} \left[\frac{4\pi^2}{g_1^2}-\frac{4\pi^2}{g_2^2}\right]=3M\left(1+\frac{3g_sK}{4\pi}\log r\right),
\eeq
From the dilaton, we see that the sum of the gauge couplings, given by (\ref{rg1}), varies as
\beq
\frac{\partial}{\partial\log\Lambda} \left[\frac{4\pi^2}{g_1^2}+\frac{4\pi^2}{g_2^2}\right]=-\frac{3K}{4}.
\eeq
The leading terms give precisely the one-loop SV $\beta$-functions.  The subleading (in $1/N$) logarithmic term incorporates some of the effects of a decreasing number of colors, but I have not found a convincing argument to explain its normalization (which may be related to a choice of renormalization scheme.)

\subsection{Five-form and Warp Factor}

The usual ansatz for the RR five-form and the geometric warp factor is
\beq 
g_s\tilde{F}_5 = d^4x \wg dh^{-1} + \star (d^4x \wg dh^{-1}) 
\eeq
which is manifestly self-dual.  $\tilde{F}_5$ satisfies a Bianchi
identity, 
\beq d\tilde{F}_5 &=& H_3 \wg F_3 = \frac{G_3 \wg
\bar{G}_3}{\bar{\tau}-\tau} \\
&=& \frac{3g_s (M\alpha')^2}{8}\left(1+\frac{15g_sK}{4\pi}\log r +\frac{g_sK}{2\pi}\log(\sin\frac{\theta_1}{2}\sin\frac{\theta_2}{2})\right)\nonumber \\
& &\qquad \times \, \,\frac{dr}{r}\wg \zeta\wg \Omega_{11}\wg \Omega_{22} + O((g_sK)^2).  
\eeq 
The warp factor $h$ which satisfies the equations of motion is (in the
standard near-horizon limit, and at linear order in $g_sK$)
\beq
h(r,\theta_1,\theta_2)=\frac{L^4}{r^4}\left(1+\frac{3g_sM^2}{2\pi N}\log r (1+\frac{3g_sK}{2\pi}(\log r+\half) +\frac{g_sK}{4\pi}\log(\sin\frac{\theta_1}{2}\sin\frac{\theta_2}{2}))\right).
\eeq
To study the cascade at leading order, we may discard the angular terms.  Then we find that the effective number of units of five-form flux is
\beq
N_{eff}(r)=N+\frac{3g_s M^2}{2\pi}(\log r +\frac{3g_s K}{2\pi}\log^2 r).
\eeq
Under the radial rescaling $r\rightarrow e^{-\frac{2\pi}{3g_s M_{eff}}}r$,
$N_{eff}$ decreases by $M_{eff}-K$ units, in agreement with the argument
from Seiberg duality (up to linear order in $g_s K$.)

\section{T-dual Type IIA Brane Configurations}

There is another class of stringy constructions of the gauge theories
studied in this paper, to which we now turn.  These models are based
on brane configurations in type IIA superstring theory \cite{Hanany},
and it was argued in \cite{Dasgupta} that the T-dual of the conifold
theory with fractional branes is just such a collection of branes.
Some features of the gauge theory are quite transparent from the
perspective of these brane constructions, so in this section we will
review some relevant results of these models, mostly following the
papers \cite{Brodie,Brunner,Park}.  Throughout this section, we work
in ten flat spacetime dimensions, and take the $x^6$ direction to be
compactified on a circle.

Let us first describe the brane construction corresponding to the
conifold gauge theory without fundamental flavors.  We place one
NS5-brane along the 012345 directions (which for conciseness we will
call an NS brane), and another NS5-brane along the 012389 directions
(which we will call an NS' brane.)  We will also place D4-branes along
the 01236 directions.  At generic positions, the D4-branes must wrap
the entire $x^6$ circle.  However, because it is possible for
D4-branes to end on NS5-branes, we may also have half-D4-branes
extending from the NS brane to the NS' brane.  Clearly, there are two
types of such half-D4-branes; moreover, two half-D4-branes of
different types may fuse to become a regular D4-brane, which is then
free to move to generic values of the 45789 coordinates.  On each
stack of half-D4-branes, there is a four-dimensional gauge theory at
low energies with gauge group $SU(N_i)$, $i=1,2$, where $N_i$ is the
number of branes in each stack.  There are also bifundamental matter
fields arising from strings connecting the two stacks of
half-D4-branes.  As a result, the matter content is precisely that of
the conifold theory described in sections 3 and 4.  By performing a
T-duality in the $x^6$ direction, the half-D4-branes become fractional
D3-branes, while the NS5-branes T-dualize into the geometry of the
conifold.

To add fundamental flavors to this model, we should add branes of
higher dimension.  The most natural choice is to add D6-branes.  For
example, we may embed a D6-brane along the 0123457 directions, and
take it to be coincident with the NS brane.  This brane configuration
preserves $\mathcal{N}=1$ supersymmetry in the gauge theory.
Moreover, we find that there will be $q,\tilde{q},Q,\tilde{Q}$ flavors
from 4-6 strings which are completely analogous to the flavors found
in Section 3.  Of course, now the T-duality along the $x^6$ direction
takes the D6-brane to a D7-brane.  The paper \cite{Brunner} also
derived the superpotential (\ref{wflavors}); they found moreover that
the couplings $g$ and $h$ in (\ref{wflavors}) are related by $g=-h$.
Note that the D6-branes split into two halves on the NS brane;
however, it is necessary for both halves to be present for RR charge
conservation.  This is analogous to the necessity of flavors with both
positive and negative chirality for cancellation of gauge anomalies;
in the IIB picture the analogous requirement is that the holomorphic
embedding equation for the D7-branes be expressible in terms of the
$z_i$ variables (rather than an equation of the form $A_1=0$, for
example.)

We can now explain (heuristically) the duality cascade in this theory
from the perspective of the IIA construction via brane creation, first
without fundamentals.  If the sizes of the stacks of half-D4-branes
are unequal, then the D4-brane tension will cause the NS and NS'
branes to bend.  At some point, the NS and NS' branes may reach equal
values of $x^6$; this corresponds to a divergence of the coupling of
the $SU(N+M)$ gauge group.  We can remove this divergence by moving
the NS' brane around the circle.  As the NS and NS' brane cross, the
stack of $N+M$ half-D4-branes shrinks to zero size and then regrows,
changing its orientation to a stack of anti-branes in the process.
However, the stack of $N$ half-D4-branes now extends more than once
around the circle; in the region with the $N+M$ anti-branes sees a
stack of $2N$ D4-branes; the branes then annihilate to give $N-M$
half-D4-branes.  Thus we have a transition taking the gauge group
$SU(N+M)\times SU(N)$ to $SU(N)\times SU(N-M)$.  

With this picture in mind, the analogous cascade with D6-branes is
easy to describe.  As the NS' brane crosses the NS brane and D6-brane,
there is an additional D4-brane created due to the D6.  Thus instead
of having $2N-(N+M)=N-M$ D4-branes, we will have $2N-(N+M)+K=N-M+K$
D4-branes.  This picture reproduces the pattern of Seiberg dualities
described in Section 4.

\section{Prospects}

In this paper we have described supergravity solutions containing
D7-branes which are dual to $\mathcal{N}=1$ supersymmetric gauge
theories with fundamental flavors.  There are many questions to be
resolved, and we shall collect some of them here.

\begin{itemize}
\item We expect that the holomorphic D7-brane configurations described
in this paper are supersymmetric, but it would be nice to explicitly
verify this claim from the standpoint of $\kappa$-symmetry on the
D7-brane worldvolume.

\item Another loose end in this paper concerns higher order
corrections to the anomalous dimensions of the $A$, $B$, and $q$
fields.  These could perhaps be analyzed in supergravity along the
lines of \cite{fkt}, or in field theory.

\item It would be interesting to study D7-brane fluctuations using the
DBI action, as was done in similar setups in \cite{Kruczenski,Sakai}.
This calculation would give a check on stability of the brane
configuration, and would also give the spectrum of mesonic operators
in the field theory.  Running the AdS/CFT correspondence in the other
direction, it might also be possible to generate the complete space of
D7-brane configurations from field theory, as was done for giant
gravitons by \cite{Beasley}.

\item The supergravity solutions given in section 4 are only valid at
intermediate values of the radial coordinate $r$, so it would be nice
to find a globally well-defined solution using F-theory.  With a
solution valid in the far infrared, it might become possible to study
phenomena associated with chiral symmetry breaking.  It would also be
interesting to study nonperturbative objects such as baryons, which
arise from D3-branes wrapped on the blown-up 3-cycle of the deformed
conifold.  Such an object would be a true baryon, made of fundamental
quarks (as opposed to dibaryons or baryon vertices with nondynamical
quarks, as have been previously considered in the literature.)

\item An F-theory solution with a compact CY four-fold would also be
interesting as an ultraviolet completion of the cascading theory with
flavors.  Warped compact versions of the Klebanov-Strassler solution
\cite{KS} have been considered in the context of possible solutions to
the hierarchy problem \cite{GKP}.  The cascade studied in this paper
suggests that warped compactifications with flavor have naturally
small gauge groups at low energies; even if the numbers of units of
flux $N$ and $M$ are both large in the ultraviolet, they both decrease
via duality cascades until they are of order $K$ in the infrared.  We
leave the elucidation of these models for future work.

\end{itemize}

\section*{Acknowledgments}

I am grateful to Igor Klebanov and Chris Herzog for collaboration in
the early stages of this project and for many useful discussions.  It
is also a pleasure to thank Chris Beasley, Sameer Murthy, and Edward
Witten for helpful conversations.  This material is based upon work
supported by the National Science Foundation Grants No.  PHY-0243680
and PHY-0140311.  Any opinions, findings, and conclusions or
recommendations expressed in this material are those of the author and
do not necessarily reflect the views of the National Science
Foundation.

\end{document}